\documentclass[preprint,number,12pt]{elsarticle}
\usepackage{threeparttable}  

\usepackage{amssymb}

\makeatletter

\newcommand{\Rmnum}[1]{\expandafter\@slowromancap\romannumeral #1@}
\makeatother

\begin{document}

\begin{frontmatter}

\title{Do cycles dissipate when subjects must choose simultaneously? }

\author{Bin Xu$^{\dag,\ddag}$}
\author{Zhijian Wang$^{\ddag}$\footnote{Corresponding author. Tel.: +86 13905815529; fax: +86 571 87176578.
E-mail addresses:  wangzj@zju.edu.cn (Z.J. Wang).}}

\address{$^{\dag}$Public Administration College, Zhejiang Gongshang University, Hangzhou, 310018, China}
\address{$^{\ddag}$Experimental Social Science Laboratory, Zhejiang University, Hangzhou, 310058, China}

\begin{abstract}
This question is raised by  Cason, Friedman and Hopkins (CFH, 2012)
  after they firstly
  found and indexed
  quantitatively the cycles in a continuous time experiment. 
To answer this question, we use the data from standard RPS experiment.  Our experiments are of the traditional setting --- in each of repeated rounds, the subjects are paired with random matching, using pure strategy and must choose simultaneously, and after each round, each subject obtains only private information. This economics environment is a decartelized and low-information one.

Using
the cycle rotation indexes (CRI, developed by CFH) method, we find, the cycles not only exist  but also persist in our experiment. Meanwhile,
the  cycles' direction are consistent with 'standard' learning models.
  That is the answer to the CHF question: Cycles do not dissipate in the simultaneously choose game.

In addtion, we discuss three questions (1) why significant cycles are uneasy to be obtained in traditional setting  experiments; (2) why CRI can be an iconic indexing-method for 'standard' evolution dynamics; and (3) where more cycles could be expected.

\end{abstract}






%
%
%
%
%
%

\end{frontmatter}



\section{Introduction}
Cycle is a significant social evolution phenomenon.
Evolutionary game theory predicts cyclic behavior,
but such behavior has rarely been reported clearly in previous experimental economics work~\cite{Friedman2012,Nowak2012}.
Only quite recently, in a continuous time laboratory environment, Cason, Friedman and Hopkins (CFH) found clear persistent cycles in population strategy space. Then,  CFH asked a question: Do cycles dissipate when subjects must choose simultaneously?~\cite{Friedman2012}

The question can be simplified and specified as: in traditional setting experimental games~\cite{Friedman1996,Friedman1998,Samuelson2002}, does the cycle exist and persist?
According to our experimental data and the method of CFH's
cycle rotation index (CRI)~\cite{Friedman2012},
Our results show that, at least in a traditional setting Rock-Paper-Scissors (RPS) game in which the
subjects must choose simultaneously,
the cycles exist, persist and do not dissipate.

In this letter, we introduce the traditional setting standard RPS experiment and the $CRI$ method at first; Then we report the cycles and the support materials; discussion and summary last.

\section{A traditional setting experiment}
In RPS game, rock ($R$) beats scissors ($S$) which beats paper ($P$) which beats rock.  If the payoff matrix is   Figure~\ref{fig:randomly matched RPS paoffmarix}(Left), the game is a standard RPS~\cite{Sandholm2011,Nowak2012}.
We use this game in our experiments to test the cycles' existence or dissipations. We use this game for two reasons. (1) It is the simplest; (2) In replicator dynamic, the real part of the eigenvalues of this game is 0 and  the endless cycles are expected.


Our laboratory experiment is of the \emph{traditional setting}~\cite{Friedman1996,Friedman1998,Samuelson2002,selten2008,Binmore2001}. In each round, every player must choose one of $(R, P, S)$. One will get 2 points if win, $1$ for tie and 0 for loss. In each of the 12 sessions\footnote{Total 72 student subjects from Zhejiang University involve in the 12 game sessions and on average, each earns 50 RMB in 1.5 to 2 hours.}, 6 subjects play the game anonymously over 300 rounds with
random matching.
 Only after   all subjects make  (privately) their own decisions, the private information (only she and her opponent's strategies and payoffs) of this round are fed back privately. In such a game setting, the subjects are controlled to choose simultaneously.

 For a given round ($t$) in this game, the social state $x(t)$ is in (and only in) one of the discrete 28 states  triagle~\cite{Nowak2012} (Figure~\ref{fig:randomly matched RPS paoffmarix}, Right). Along rounds, the state should transit and form a trajectory~\cite{Friedman2012,Binmore2001} in which cycle is expected by many 'standard' models~\cite{Sandholm2011,Friedman2012}.

\section{Cycle Rotation Indexes}
CRI (Cycle Rotation Indexes) is developed by CHF~\cite{Friedman2012}.
To calculate this index, they first constructed a
line segment in the projection of the simplex into the two-dimensional plane between the Nash equilibrium and the simplex edge, illustrated as the 
 line segment $S$ in Figure~\ref{fig:randomly matched RPS paoffmarix}.
This line segment $S$
serves as a Poincare section for cycle counting.

Specifically, in a given time interval [$t_0,t_1$], how many times (CCT, counterclockwise transits, $C^+_{t_0,t_1}$) the social state crosses $S$ from left to right and how many times
(CT, clockwise transits, $C^-_{t_0,t_1}$) from right to left
can be counted. Then, the cycle rotation indexes (denoted as $\theta$)  can be written in formula as
\begin{equation}\label{eqcri}
     \theta = \frac{C^+ - C^-}{C^+ + C^-}.
\end{equation}
Thus the CRI $\theta$ values near 1 (-1) indicate  consistent counter-clockwise (clockwise) cycles, and values near 0 indicate no consistent cycles. We use the subscript $(t_0,t_1)$ to denote the time interval of the observation. For example, $\theta_{t_0,t_1}$ is the result of $\theta$ calculated in the trajectory $x(t)$ from round-$t_0$ to round-$t_1$ including $t_1$-$t_0$ times transitions in the strategy space. Mathematically, CRI is of time odd ($\theta_{t_1,t_2}$=-$\theta_{t_2,t_1}$).

In this way, the question --- whether cycles dissipate --- becomes testable: the null hypothesis is $H_0^d$: $\theta = 0$. If $H_0^d$ can not be rejected in the data, cycles do dissipate; Alternatively, cycles do not dissipate.

\section{Results}
\begin{figure}
\centering%
\includegraphics[angle=0,width=3cm]{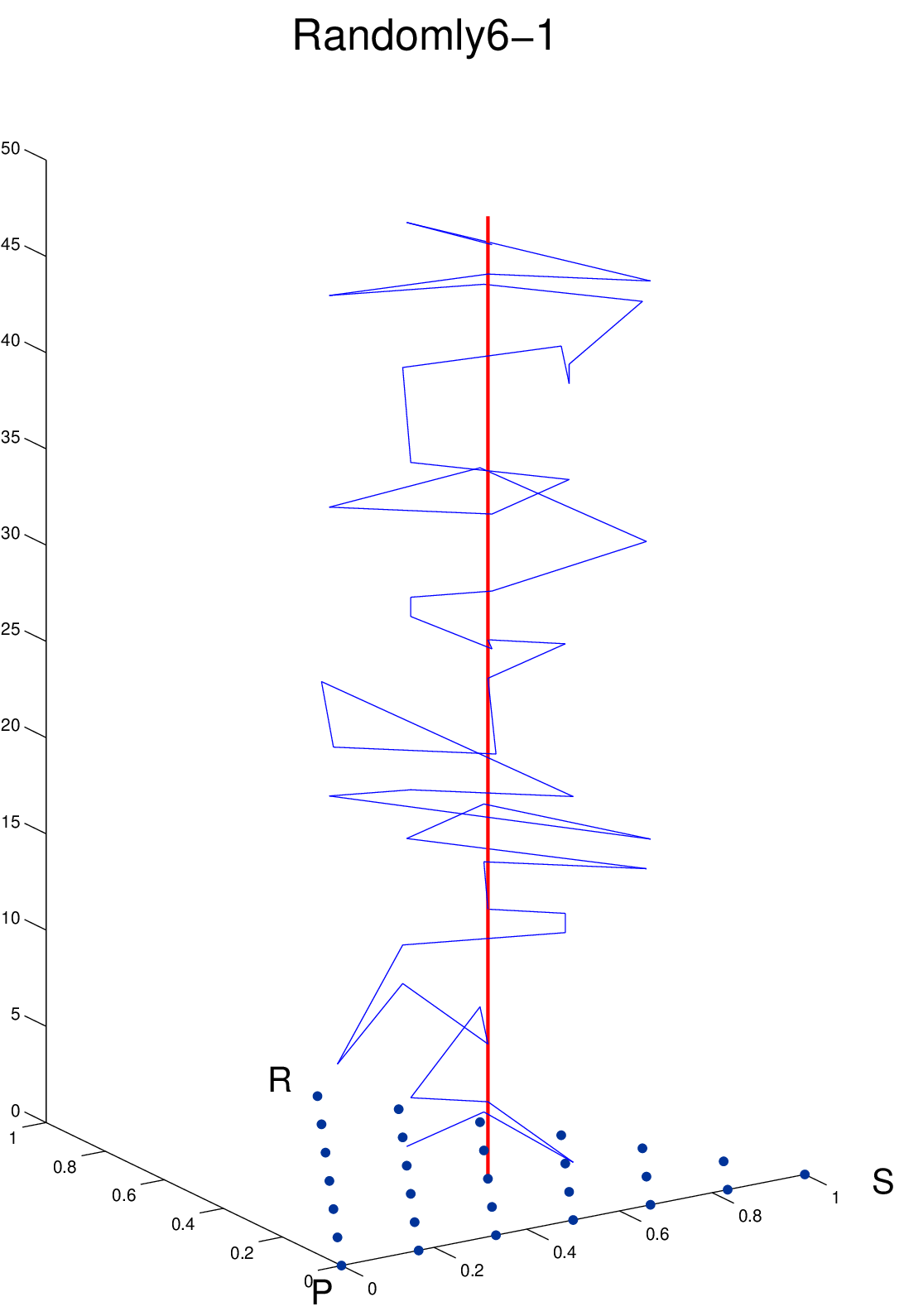}
\includegraphics[angle=0,width=4.5cm]{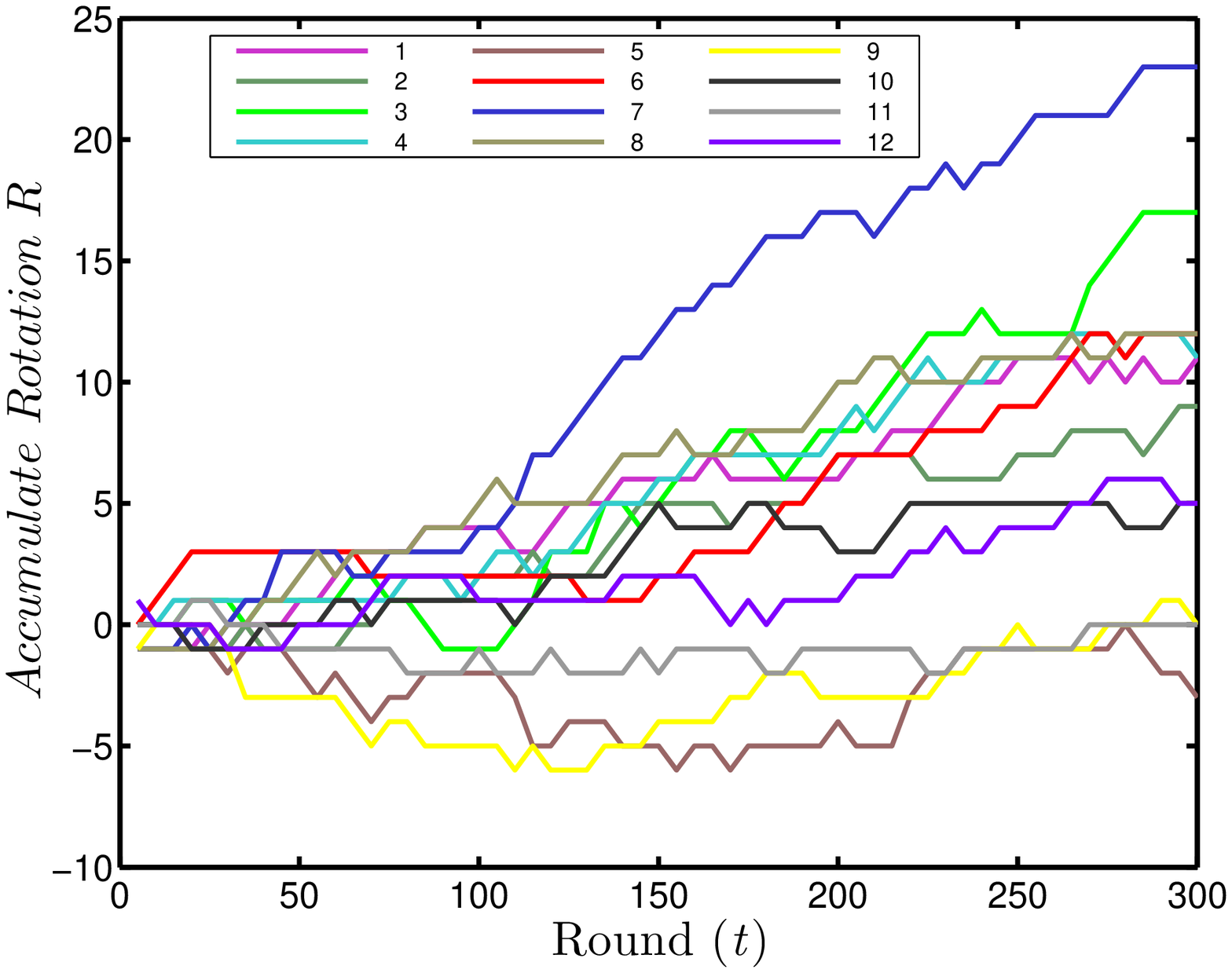}
\caption{(Left) Trajectory. Data from the 1$st$ experimental session. The vertical axis represents
 round ($t$) from 1-$st$ to 50-$th$.
 (Right) Accumulated rotation $R_{1,t_1}$ (:= $C^+_{1,t_1}$-$C^-_{1,t_1}$) of the  12 sessions.
\label{fig:Spring+AccR}}
\end{figure}

We  collect data from the 12 sessions each having 300 rounds. Figure~\ref{fig:Spring+AccR} is an evolution trajectory. Mean observation of the RPS, denoted as $\bar{x}_{1,300}$, is (0.357,0.321,0.322) which is close to Nash equilibrium  \emph{NE}  ($\frac{1}{3},\frac{1}{3},\frac{1}{3}$). Meanwhile, the observed distributions  satisfied the distributions predicted by multi-normal distributions  (the principle of Maximun entropy~\cite{xuetal2012Maxent})~\footnote{$\chi^2$  goodness fit testing, $p<0.05$ in the 12 sessions, respectively.}. This results are similar to the results called as the cycles amplitudes in~\cite{Friedman2012} and called as the distances from the center in~\cite{Nowak2012}.

\begin{figure}
\centering
\includegraphics[angle=0,width=3cm]{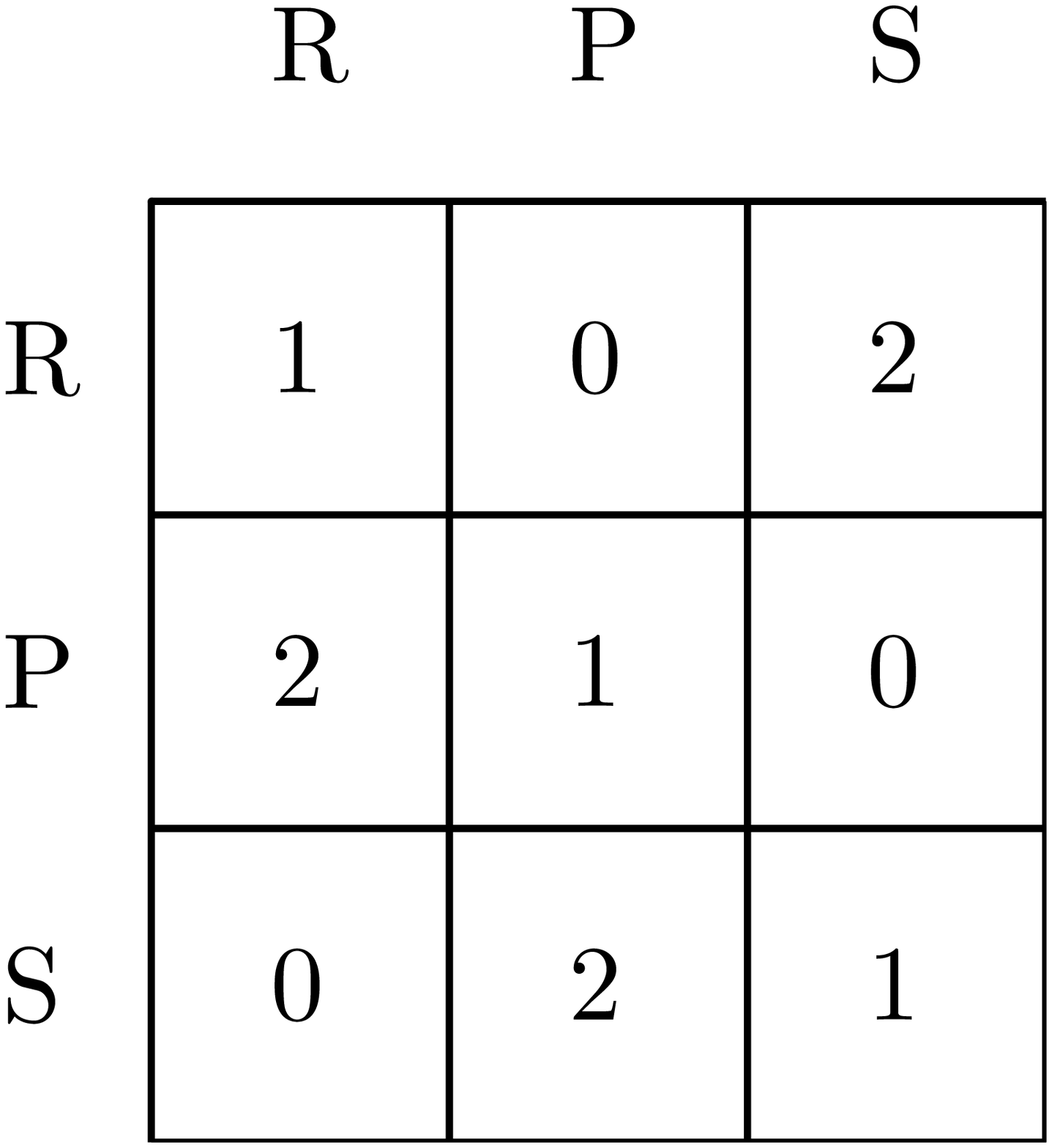}  ~~~~~~~~
\includegraphics[angle=0,width=4cm]{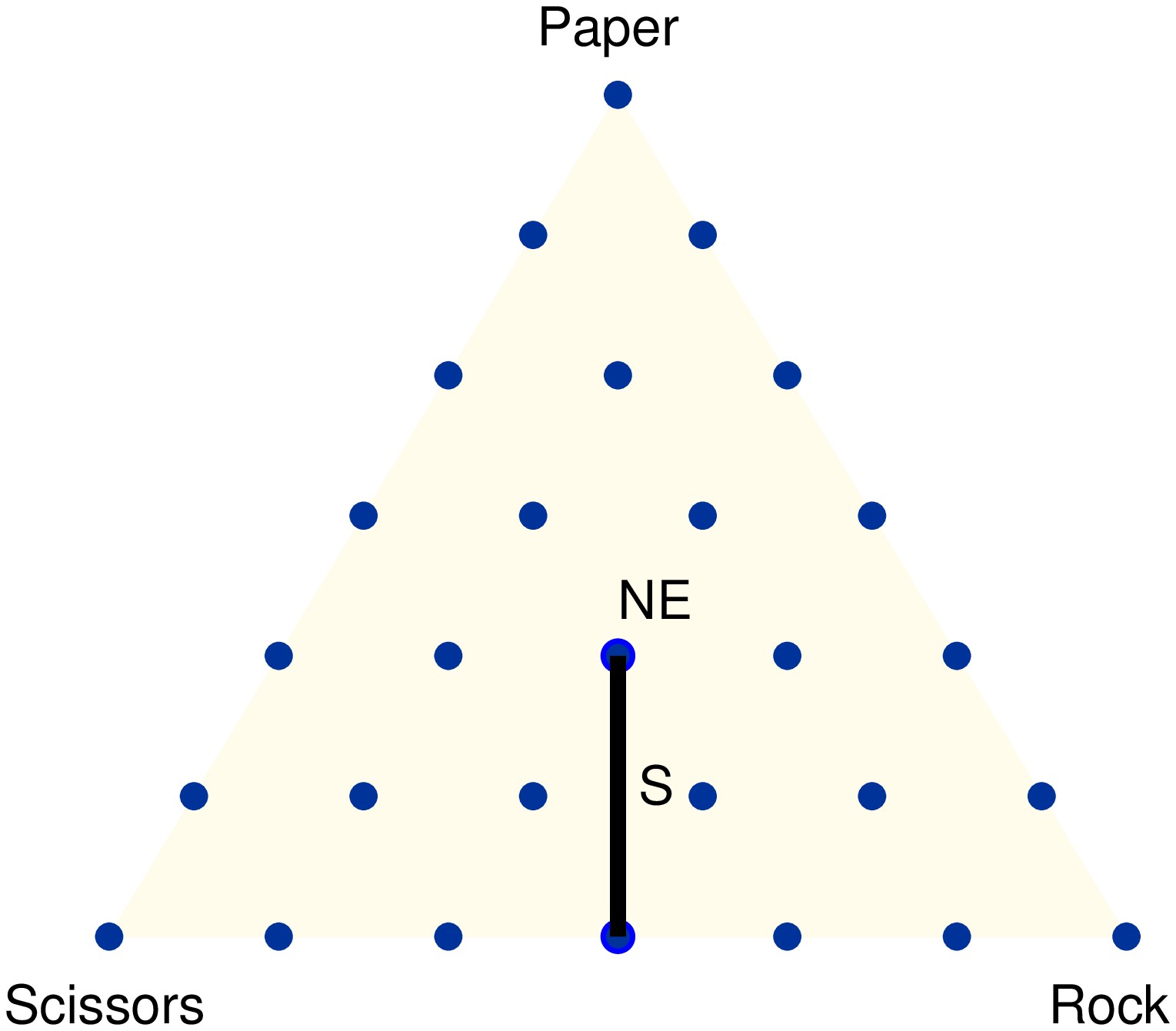} ~~~~~~~~
\caption{(Left) Payoff Matrix of standard RPS Game. (Middle) Social state space of the Game. Each of the 28 dots denotes an observable states. The state denoted as $(a,b)$ is the state in which proportion of the $R$ ($P$) strategy in the population is  $a$ ($b$).
  The segment $S$ is the Poincare section for $C^\pm$ counting. \emph{NE} appears as the reference is the centroid of the simplex at
($\frac{1}{3}$, $\frac{1}{3}$) in the game.
\label{fig:randomly matched RPS paoffmarix}}
\end{figure}

 Mainly we use CRI
 to report  Rotation strength and direction, then the evidences of  the cycles' existence and  the persistence.
 In this report, we set the the segment  from \emph{NE}
 to the low edge of the triangle as the Poincare section\footnote{
 For simple,  the segment in Figure~\ref{fig:randomly matched RPS paoffmarix}(Right) is shifted (to left-down in 10$^{-3}$) slightly
 and  does not pass though any of the 28-state.}.

\begin{figure}
\centering%
\includegraphics[angle=0,width=4cm]{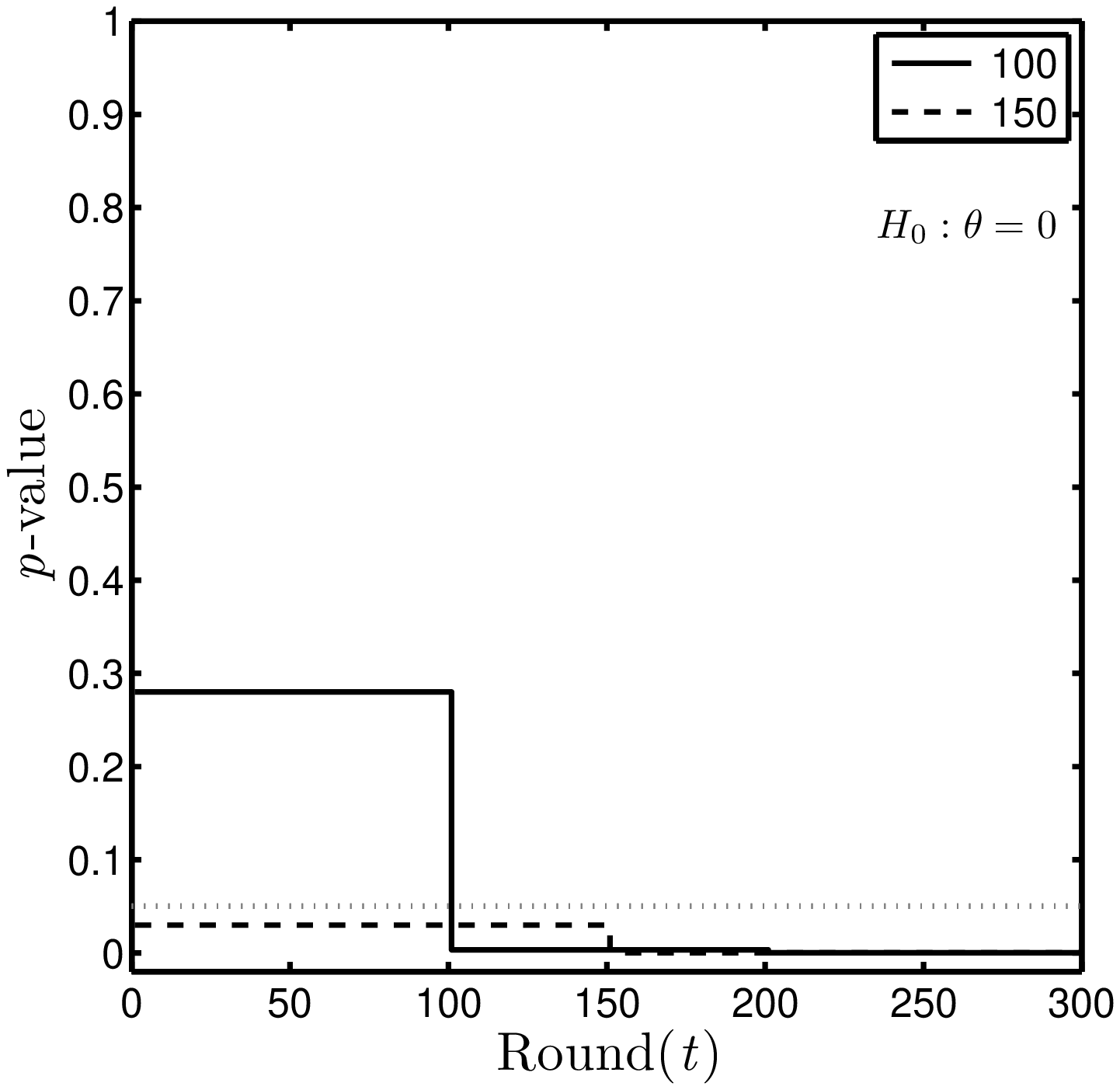}
\includegraphics[angle=0,width=4.5cm]{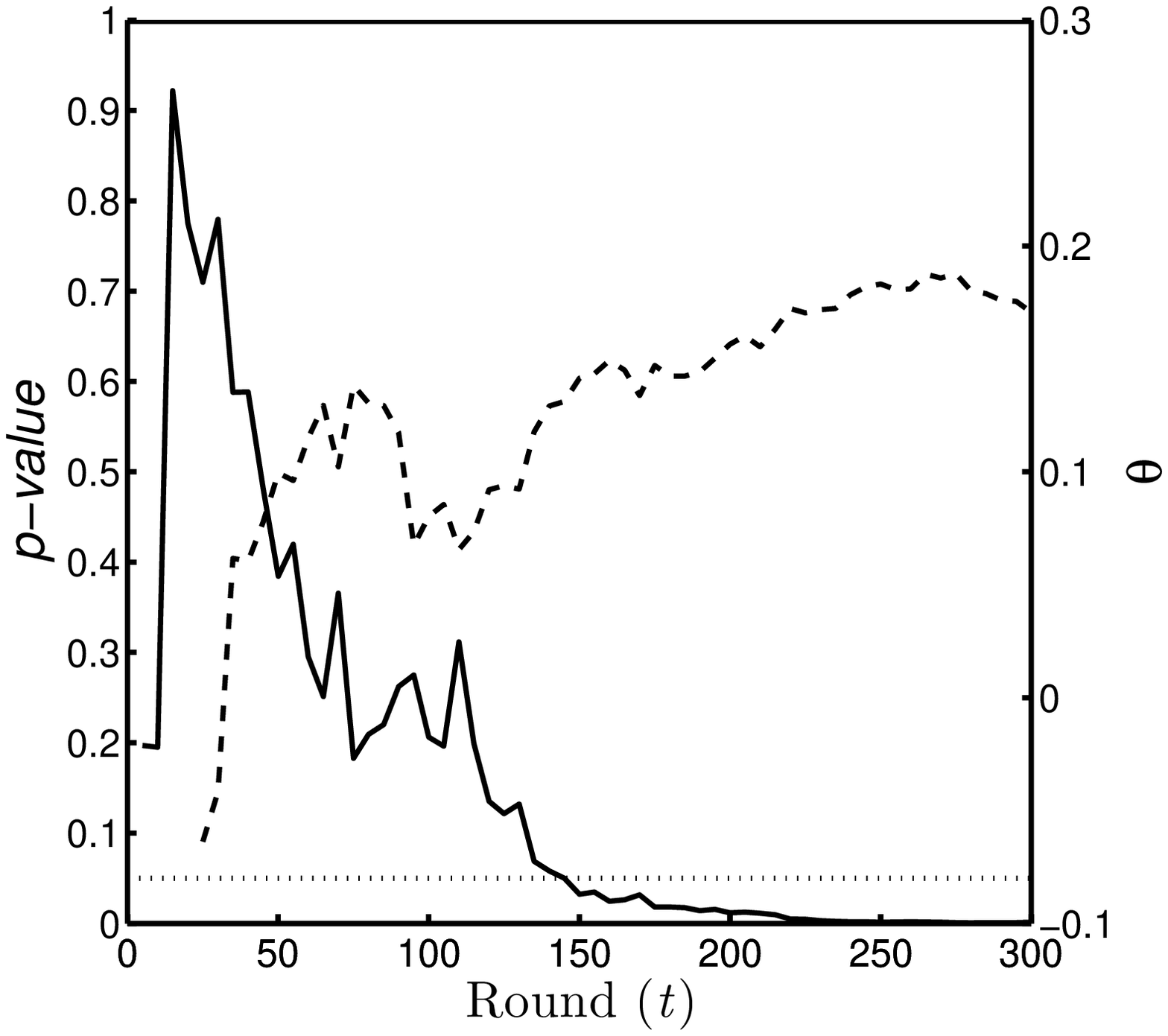}
\caption{The $p$-values of $t$-$test$ for CRI ($\theta$) with $H_0^d$: $\theta$=0.
(Left) For different sample time interval (100 and 150, refer to Table~\ref{tab:c+c-theta}); (Right) Increasing $t_1$ for $\theta_{1,t_1}$ testing as $t_1$ = 5, 10, 15, ..., 300 (in solid) and the observed $\theta_{1,t_1}$  (in dash).
\label{fig:pValue}}
\end{figure}

 \textbf{Result-1}  To estimate the strength of the rotations,  we calculate accumulated rotation $R_{1,t_1}$ ($R_{1,t_1}$:= $C^+_{1,t_1}$-$C^-_{1,t_1}$) and $\theta_{1,t_1}$ of each  session. The $R_{1,t_1}$ results of the 12 sessions are exhibited in Figure~\ref{fig:Spring+AccR} (Right). The $\theta_{1,t_1}$ is in dash line in  Figure~\ref{fig:pValue} (Right).  In the 12 samples, $\bar{R}_{1,300}$=8.50$\pm$2.17 ($ > 3\sigma$) and $\bar{\theta}_{1,300}$=0.17$\pm$0.04 ($ > 4\sigma$).


\textbf{Result-2} To estimate the existence of the cycles, we use $C^\pm_{1,300}$ and $\theta_{1,300}$
(2-4 columns in Table.~\ref{tab:c+c-theta}).
 Statistical results are (1) $H_0^d$: $\theta_{1,300}$=0 can be rejected ($p=0.016<0.05$, $t$-$test$, two-tailed, $d.f.$=11). Meanwhile, $H_a$: $\bar{\theta}_{1,300}<$0 can be rejected ($p$=0.0083, $t$-$test$, one-tailed, $d.f.$=11). (2) equal hypothesis (the two samples paired $t$-$test$ of the 12 paired $C^+$ and $C^-$) can also be rejected ($p=0.017<0.05$,  $t$-$test$, two-tailed, $d.f.$=11). So, the dissipation of cycles in the data can be rejected; Alternatively, the cycles exist.

  \textbf{Result-3} To estimate the persistence of the cycles,  we use the $p$-values (1) from $t$-$test$ of $H_0^d$: $\theta$=0; The results are shown in Figure~\ref{fig:pValue} (Left) and Table.~\ref{tab:c+c-theta}; (2) from $H_0^d$: $\theta_{1,t_1}$=0, the $p$-values are shown in Figure~\ref{fig:pValue} (Right) as a function of $t_1$.   The $p$-values are below 0.05 when rounds $t>$100. So, in the game, the persistence of the cycles is supported by the data.

 \textbf{Result-4} For cycles' direction, we use the materials in \emph{Result-2}. $\bar{\theta}_{1,300}>0$  means the cycles are of counter-clock-wise, which meets the expectations of evolution dynamics (see Figure~\ref{fig:RpsBNN}).

According to the sub-results above, in summary, the hypothesis of cycle dissipation ($H_0^d$) can be rejected (at least at $3\sigma$ level).

\begin{table}[htbp2]
\centering
\tiny
\begin{threeparttable}
\caption{\label{tab:c+c-theta} CRI ($\theta$) in the standard RPS game}
\begin{tabular}{|c|ccr|rrr|rr|}
  \hline
   \hline
    Session	& $C^+$	& $C^-$		& $\theta_{1,300}$		&$\theta_{1,100}$	&$\theta_{101,200}$	 &$\theta_{201,300}$		 &$\theta_{1,150}$	&$\theta_{151,300}$	\\
    \hline											
1	&17	&28	&0.24	&0.22	&0.20	&0.29	&0.25	&0.24	\\
2	&16	&25	&0.22	&0.08	&0.38	&0.14	&0.22	&0.22	\\
3	&24	&41	&0.26	&-0.07	&0.39	&0.33	&0.2	&0.3	\\
4	&16	&27	&0.26	&0.20	&0.38	&0.14	&0.43	&0.17	\\
5	&25	&22	&-0.06	&-0.11	&-0.14	&0.07	&-0.2	&0.09	\\
6	&13	&25	&0.32	&0.20	&0.29	&0.38	&0.11	&0.50	\\
7	&22	&45	&0.34	&0.18	&0.48	&0.33	&0.33	&0.35	\\
8	&15	&27	&0.29	&0.33	&0.38	&0.14	&0.33	&0.24	\\
9	&20	&20	&0.00	&-0.38	&0.13	&0.27	&-0.17	&0.25	\\
10	&26	&31	&0.09	&0.11	&0.10	&0.13	&0.15	&0.00	\\
11	&26	&26	&0.00	&-0.05	&0.00	&0.07	&-0.03	&0.09	\\
12	&23	&28	&0.10	&0.05	&0.00	&0.25	&0.07	&0.13	\\
\hline
$Mean$ 	&20.25	&28.75	&0.17	&0.06	&0.22	&0.21	&0.14	&0.22	\\
$S.E.$	&1.35	&2.1	&0.04	&0.06	&0.06	&0.03	&0.06	&0.04	\\
\hline
\end{tabular}
\begin{flushleft}
\end{flushleft}
\end{threeparttable}
\end{table}

\section{Discussion}

In a standard RPS experiment with traditional setting --- multi-subjects, randomly paired, multi-rounds repeated, each subject using pure strategy and making decision simultaneously with limited information
~\cite{Friedman1996,Friedman1998,Samuelson2002,selten2008,Binmore2001}, our answer to the question is: the cycles do not dissipate.

We analyze the three investigations (CFH~\cite{Friedman2012} and this letter) by answering two questions (1) why CRI is important; (2) why the cycle is uneasy to be detected in traditional setting? (3) where more cycles could be expected?

\subsection{Why CRI is important}

Cycle is a significant social evolution phenomenon and is seeked for long in experimental games~\cite{Binmore2001,Friedman1996,benaim2009learning}. With CRI~\cite{Friedman2012}, the cycle can be indexed quantitatively. The importance can be interpreted briefly as following.

\begin{figure}
\centering%
\includegraphics[angle=0,width=4cm]{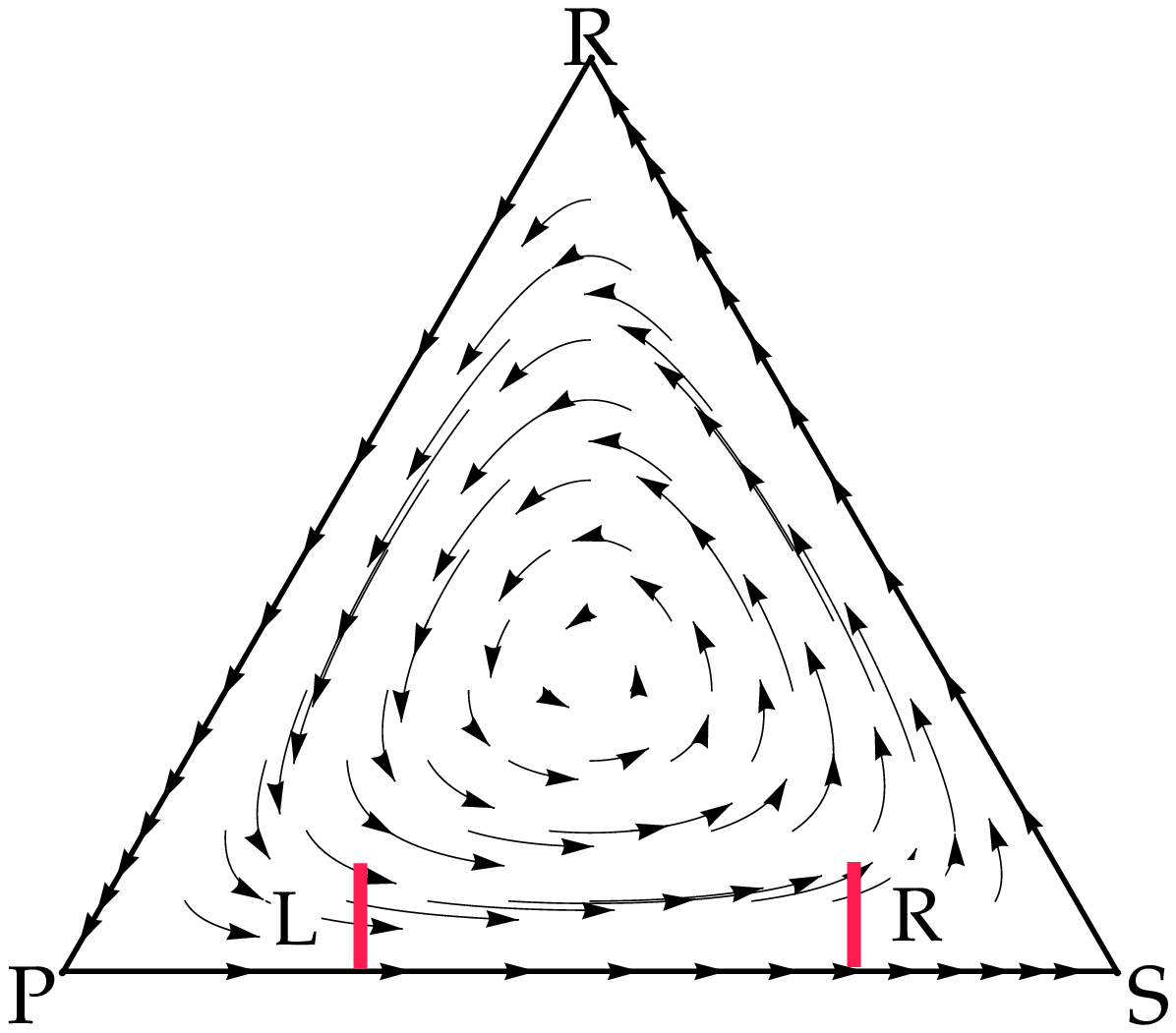}
\includegraphics[angle=0,width=4cm]{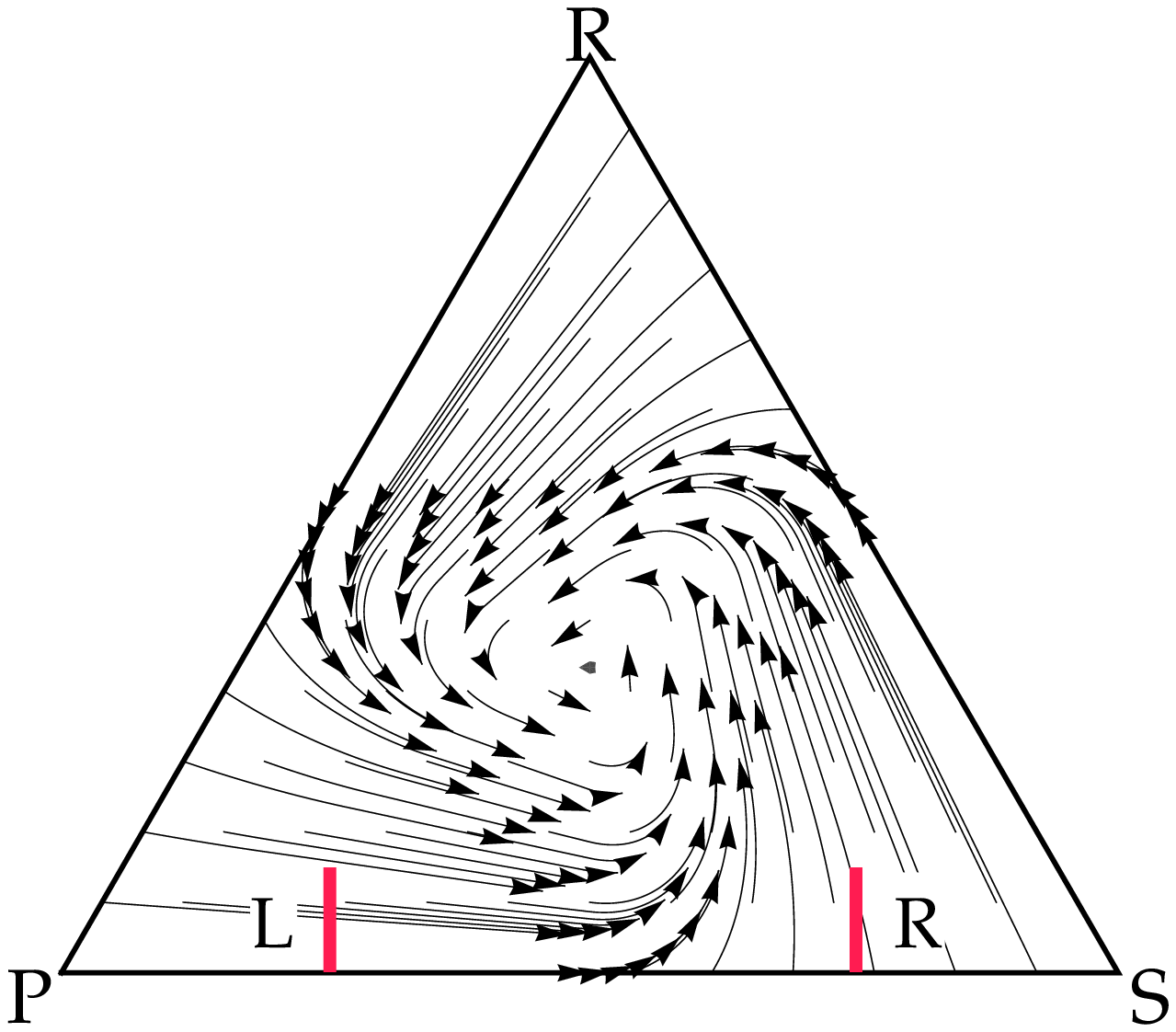}
\caption{Evaluating dynamics models with CRI  in standard RPS game. The counter-clock-wise cycles prediction using Dynamo~\cite{Dynamo2012} of RD and BNND. $L$ and $R$ are of the Poincare  sections. For the two $\theta$-values in $L$ and $R$, RD predicts equal but BNND  unequal.
So, RD and BNND  can be competitive and testable with CRI.
\label{fig:RpsBNN}}
\end{figure}

There are many 'standard' evolution dynamics, e.g., replicator dynamics (RD), logit dynamics (and other smoothed best reply dynamics), Brown-von Neauman-Nash dynamics (BNND) and so on~\cite{Sandholm2011}. The predictions from the  dynamics differ and is rarely tested by experiment quantitatively. CRI can serve  for this aim.
Using CRI, the models can be testable experimentally now. For example,
Figure~\ref{fig:RpsBNN} explains a probable way using CRI method to distinguish the two models, RD and BNND, to see who meets standard RPS experiments better.

The variable of CRI is time odd ($\theta_{t_1,t_2}$=-$\theta_{t_2,t_1}$).
Meanwhile, the mean observation ($\bar{x}_{t_1,t_2}$=$\bar{x}_{t_2,t_1}$) is time even.
Time even (odd) means the observable should be invariant (opposite) under time reversal transformation.
CRI can capture the evolution probability current in the Poincare section in long run view.
In experimental game data, similar observable e.g. velocity field~\cite{XuWang2011ICCS} \cite{Xuetal2011} and angular momentum ~\cite{WangXu2012} are also time odd and also indicate  the existence of dynamics pattern, but the methods seem less straightforward. The consistence of the methods needs further investigations.
We believe the
CRI could be a variable, as important as the experimental mean observation $\bar{x}$ for  testing  classical equilibrium theorems,  for evolution theorems testing.

\subsection{Why the cycles are uneasy to be detected in traditional setting RPS game}
Cycles are uneasy to be detected~\cite{benaim2009learning}.  What is the reason? Besides the CRI method, we suggest that the sample size (number of the rounds and sessions) could be the main reason. 
According to the \textbf{result-3} and Figure~\ref{fig:pValue}(Right) from our data, it can be noticed that
 CRI $\theta_{1,t_1} > 0$ could be significant only when $t_1 > 150$.

These mean that, to obtain cycles in the RPS game, larger numbers of the experimental rounds are necessary. In $Z$-Tree\cite{zTree2007}-like softwares using experiments,   the number of rounds usually is less than 150 (e.g.~\cite{Friedman1996}), from which to detect cycles should be uneasy.

%

%

In economics, as suggested by Hayek~\cite{Hayek1945}, one of the most interesting phenomena is that, even though everyone has only limited and localized information, global social order can form. Traditional setting game is in the decentralized  economics environment. 
The experimental finding of the cycles in the traditional setting game suggests that the dynamics orders --- cycles --- \emph{exist} in laboratory even though long time is needed. 


\subsection{Where more cycles could be expected}

Thanks to the wide parameters of the CFH's experiments, we can begin to compare cycles in variety  environments.
For the existence of cycles, a three dimensional (information, strategy and time) analysis is demonstrated in Table~\ref{3ExpCompare} in which CHF, HSNG and WX presents the experiments reports~\cite{Friedman2012}, ~\cite{Nowak2012} and this letter.
  %
The results indicate that  none of the three dimensions could be a unique necessary condition for the existence of cycles. 

 There are two straightforward puzzles when compare the two reports~\cite{Friedman2012,Nowak2012}. (1) In the standard (called as neutral in~\cite{Nowak2012}) RPS games,  why no cycles is reported in the entire information~\cite{Nowak2012}? But the cycles exist in our limited information condition. (2) The Bad-RPS games in~\cite{Nowak2012} is exactly same as the CHF's $U_a$-DP (see Table~\ref{3ExpCompare}), why the Bad-RPS reports no evidence of cycles? Even more, extended questions can be, e.g., can cycles be  found out in mixed strategy 2$\times$2 games? The cyclic velocity vectors field obtained in the experimental coyness and Philandering games~\cite{XuWang2011ICCS} strongly suggests the existence of the cycles in 2$\times$2 games. 


\begin{table}[htbp2]
\centering
\tiny
\begin{threeparttable}
\caption{\label{3ExpCompare} When cycles exist in RPS-like games}
\begin{tabular}{|c|c|c|c|c|c|c|c||c|c|}
   \hline		
&Game	& Info.&	Strategy&Time&Cycles	  \\
\hline
&	&	Entire(+)&   Pure(+)	&	Continue(+) &	Exist(+)            \\
&	Stable    &	Private(-)&    Mixed(-) &	Discrete(-) &   Dissipate(-)   \\
\hline
HSNG (G)    &	S &		+&		+&	- & \\
HSNG (N)    &	N &		+&		+&	- & \\
HSNG (B)    &	U &		+&		+&	- & \\
CHF (C)&	S&		+&	-&		+&	+ 	\\
CHF (C)&	U&		+&	-&		+&	+ 	\\
CHF (DM)&	S&		+&	-&	-&	- 	 \\
CHF (DP)&	S&		+&	+	&	-&	- \\
CHF (DM)&	U&		+&	-&	-&	+	  \\
CHF (DP)&	U&		+&	+	&	-&	+	  \\
XW  &		N &	-&		+&	-&		+ \\
\hline
\end{tabular}
\begin{flushleft}
Column-2: The stability of the dynamics includes 3 conditions: stable(S), neutral(N) and unstable(U). Column-3 to 5 are  the three dimensions: (1) Information  provided to subjects in game. It includes two conditions: the  entire population's (+) or only the private (-). (2) Options of strategy, pure (+) or mixed (-), in the (R,P,S) strategy set. (3) Option of the time for decision making, continuous (+) or discrete (-). Column-6: existence of cycle, yes (+) and no (-).
\end{flushleft}
\end{threeparttable}
\end{table}

\subsection{Summary}
In the traditional setting experiments where subjects must choose simultaneously with limited information, we find, the cycles do not dissipate.  This is our answer to the CHF's question~\cite{Friedman2012}.
We wish this report could serve as a supplement for the CHF's remarkable indexing of cycles.
All these state that \emph{there is much more to be learned about cycling}!~\footnote{We thanks Ken Binmore for the suggestions. Thanks also to Wang Shuang for the English writing corrections. Any error belongs to our own.  }

\bibliographystyle{plain}







\end{document}